\documentclass{article}
\usepackage{spconf}

\usepackage[colorlinks]{hyperref}
\hypersetup{
    linkcolor=blue,
    filecolor=magenta,      
    urlcolor=teal,
    citecolor=magenta
    }

\usepackage{multirow}
\usepackage{graphicx}
\usepackage{subfigure}
\usepackage{url}
\usepackage{amsfonts}
\usepackage{amssymb}
\usepackage{amsmath}
\usepackage{bm} 
\usepackage{algorithm}
\usepackage{algorithmic}
\usepackage[table]{xcolor}

\usepackage{enumitem}
\newcommand{\cmmnt}[1]{\ignorespaces}

\definecolor{Mygreen}{rgb}{0,0.7,0.2}

\DeclareMathOperator{\Diag}{\mathbf{diag}}

\title{A Hypothesis Testing Approach to Nonstationary Source Separation}

\name{Reza~Sameni\thanks{This work has been partially funded by the European Research Council Advanced Grant Number 320684, on Challenges in the Extraction and Separation of Sources (CHESS).} and Christian~Jutten}
\address{Department of Biomedical Informatics, Emory University School of Medicine, GA 30322, USA \\ GIPSA-lab, Universit\'e Grenoble Alpes, CNRS, Grenoble INP, 38000 Grenoble, France\\ Emails: rsameni@dbmi.emory.edu, christian.jutten@gipsa-lab.grenoble-inp.fr}

\begin{document}

\ninept
\maketitle
\begin{abstract}
The extraction of nonstationary signals from blind and semi-blind multivariate observations is a recurrent problem. Numerous algorithms have been developed for this problem, which are based on the exact or approximate joint diagonalization of second or higher order cumulant matrices/tensors of multichannel data. While a great body of research has been dedicated to joint diagonalization algorithms, the selection of the diagonalized matrix/tensor set remains highly problem-specific. Herein, various methods for nonstationarity identification are reviewed and a new general framework based on hypothesis testing is proposed, which results in a classification/clustering perspective to semi-blind source separation of nonstationary components. The proposed method is applied to noninvasive fetal ECG extraction, as case study.  
\end{abstract}
\section{Introduction}
\label{sec:introduction}
A broad class of blind and semi-blind source separation (BSS) algorithms are based on the joint (approximate) diagonalization of a set of matrices/tensors containing second or higher order statistics of multichannel data. While a great body of research has focused on efficient methods and algorithms for approximate joint diagonalization (AJD) of a set of matrices \cite[Chs. 5 and 7]{ComonJutten2010handbook}, the original selection of these matrices remains a highly problem-specific and nontrivial issue. The problem is more critical for mixtures of nonstationary or sparse signals, for which the desired signals of interest are difficult to detect and are considered as outliers that do not necessarily influence the average statistics captured by the diagonalized matrices/tensors. In fact, many BSS algorithms perform source separation based on the average statistics from the entire dataset and are incapable to identify local variations in the signal statistics that do not have significant effects on the global statistics. 

Herein, we review some of the existing and new methods for supervised and unsupervised nonstationary sample identification (and clustering) during the formation of the matrices used in (approximate) joint diagonalization. As a result, the diagonalized matrices are calculated more targeted (and prior-based), making them more appropriate for nonstationary source signals. The idea is shown to be of great interest for unsupervised and semi-supervised source separation, and closely related to the problem of clustering and classification of samples in imbalanced and nonstationary scenarios.

In Section \ref{sec:nonstationaritydetection}, we review methods for nonstationarity detection. These methods are later used in Section \ref{sec:NSCA}, where we demonstrate how nonstationarity can be used as a means of source separation. As proof of concept, in Section \ref{sec:casestudy}, the proposed method is applied to the problem of noninvasive fetal electrocardiogram extraction.

\section{Nonstationarity Identification}
\label{sec:nonstationaritydetection}
Nonstationarity generally refers to signal attribute variations in time, space, frequency, time-frequency, stochastic properties, etc. In the sequel, we review the concept of nonstationary in different domains and how it can be quantitatively identified. Our data model is
\begin{equation}
\bm{x}_k = \mathbf{A}_k \bm{s}_k,
\label{eq:datamodel}
\end{equation}
where $k$ denotes time, $\bm{x}_k \in \mathbb{R}^n$ are multivariate observations, $\mathbf{A}_k$ is a linear invertible matrix and $\bm{s}_k\in \mathbb{R}^n$ is the source vector.



\subsection{Density Function Variations}
\label{sec:pdfvar}
The probability density function (PDF) of the observations in (\ref{eq:datamodel}) is 
\begin{equation}
f_X(\bm{x}_k;k) = \frac{1}{|\det \mathbf{A}_k|} f_S(\mathbf{A}_k^{-1}\bm{x}_k;k)
\end{equation}
which is a function of the source vector density function and the (generally time-variant) mixing matrix. Apparently, the temporal nonstationarity of the source signals or the mixing matrix result in local variations of $f_X(\cdot;k)$. Now suppose that regardless of the local variations due to nonstationarity one fits an empirical PDF $\hat{f}_X(\bm{x})$ over the observed samples. Noting that $\hat{f}_X(\bm{x})$, or equivalently the empirical cumulative distribution function (CDF) $\hat{F}_X(\bm{x})$, are no longer functions of time ($k$), one can conduct a sample-wise \textit{hypothesis test} for nonstationarity detection, to identify the ``odd events'' that locally differ from the global density function $\hat{f}_X(\bm{x})$:
\begin{equation}
\begin{array}{rl}
         \mathcal{H}_0: \bm{x}_k \nsim \hat{f}_X(\bm{x}) \\
         \mathcal{H}_1: \bm{x}_k \sim \hat{f}_X(\bm{x})
    \end{array}
\label{eq:hypothesistestGeneral}
\end{equation}
Various statistical tests exist for evaluating whether or not a given data sample is drawn from a given PDF. The Cram\'er--von Mises, the Anderson--Darling test and their modified versions are among the classical tests \cite[Ch. 4]{d1986goodness}, \cite[Ch. 6.3]{kvam2007nonparametric}. 

Considering that the observations are commonly band-limited signals that satisfy the Nyquist rate, a nonstationary event is not an isolated point, but rather consists of a series of successive samples. Now suppose that we want to assess whether a segment of $p$ successive data points $\{\bm{x}_{k-p+1}, \ldots, \bm{x}_k\}$ corresponds to a nonstationary event, i.e., the null hypothesis in (\ref{eq:hypothesistestGeneral}) holds for these samples. In an Anderson--Darling test the following statistics is calculated:
\begin{equation}
A_k^2 = -p -\frac{1}{p} \textstyle\sum_{i=1}^p (2i - 1)\log[\hat{F}_X(\bm{z}_i)(1 - \hat{F}_X(\bm{z}_{p+1-i}))]
\end{equation}
where $\{\bm{z}_{1}, \ldots, \bm{z}_p\}$ is the sorted version of $\{\bm{x}_{k-p+1}, \ldots, \bm{x}_k\}$, in non-decreasing order. Depending on the segment length $p$, $A_k^2$ can be checked versus tabulated values to accept or reject the null hypothesis with a desired confidence (p-value) \cite[Ch. 4]{d1986goodness}. As a continuous index, $A_k^2$ can be updated over time and compared with desired thresholds to identify the nonstationary points.

\subsection{Reference-Based Nonstationary Identification}
A popular approach for nonstationarity identification is to use a reference channel to solve the hypothesis test in (\ref{eq:hypothesistestGeneral}). Various forms of this approach have been used in the literature, mostly without explicit reference to a hypothesis test. 

In \cite{Sameni2008a}, the R-peaks of electrocardiogram (ECG) signals were detected from a reference channel and used as a means of deriving variable periods of the ECG (called the cardiac phase signal), which were
used for extracting the ECG from noise by periodic component analysis. In \cite{Amini08}, a reference lead was used to identify the artifacts due to a magnetic resonance device that obscured simultaneously acquired EEG signals. In \cite{SameniGouyPailler2014}, electrooculogram (EOG) artifacts of the electroencephalogram (EEG) were identified by thresholding the energy envelope of an EOG reference lead, which was closely placed to the eye. In \cite{jamshidian2019temporally}, thoracic electrodes were used to identify the R-peaks of maternal ECG from the abdominal leads used for fetal ECG extraction. In \cite{samadi2011identification,samadi2013reference}, a reference-based method was used for defining periods including interictal epileptiform discharges for identification of brain regions involved in a reference state from intracerebral EEG.


\subsection{Observations Moments/Cumulants Tracking}
The \textit{moment generating function} of $\bm{x}_k$ defined as \cite[Ch. 2]{Hyvarinen2001}:
\begin{equation}
    \phi_X(\bm{\omega};k) = \mathbb{E}\{\exp(j\bm{\omega}^T\bm{x}_k)\}
\end{equation}
uniquely characterizes the distribution of the random vector $\bm{x}_k$. Using the Taylor expansion of $\exp(\cdot)$, variations of the PDF can be associated to and tracked by monitoring the first (mean), second (covariance), and higher-order moments and cumulants of the observations. This gives an alternative solution for solving the hypothesis test proposed in Section \ref{sec:pdfvar}. Accordingly, depending on the source of nonstationarity, one can estimate and track the first few cumulants of the observations over sliding windows, to identify the nonstationary events.


\subsection{Adaptive Source Separation}
Adaptive source separation algorithms are well-known in the literature. As proof of concept, let us consider the \textit{equivariant adaptive source separation via independence} (EASI) \cite{Cardoso1996}. In this method, the separating matrix, here denoted by $\mathbf{W}_k$, is adaptively updated using an equivariant serial update:
\begin{equation}
\mathbf{W}_{k+1} = \mathbf{W}_{k} - \lambda_k \mathbf{H}(\bm{y}_k) \mathbf{W}_{k}
\label{eq:EASI}
\end{equation}
where $\bm{y}_k = \mathbf{W}_{k} \bm{x}_k$ are adaptive estimates of the independent sources and $\mathbf{H}(\cdot)$ is a nonlinear function of the estimated sources cumulants. For time-varying mixtures, the algorithm seeks the demixing matrix such that in steady state $\mathbf{W}_{k}\mathbf{A}_{k}$ approaches identity and $\|\mathbf{H}(\bm{y}_k)\|\rightarrow 0$. Now suppose that the mixing matrix $\mathbf{A}_{k}=\mathbf{A}$ is constant, but the sources are nonstationary. As a result, the function $\mathbf{H}(\cdot)$ deviates from zero as the recursion reaches the nonstationary epochs. The nonstationary points can therefore by detected by thresholding the norm of this measure over time. A similar approach is applicable to other stochastic gradient ascent/descent algorithms for independent component analysis (ICA), which use sample-wise updates instead of global averaging \cite[Ch. 3.2]{Hyvarinen2001}, \cite[Ch. 4.5]{ComonJutten2010handbook}.  

\subsection{Spectral Nonstationarity}
\label{sec:spectral}
Spectral variations are another common type of nonstationarity \cite{yeredor2010second}, which can be identified by using time-frequency signatures of the nonstationary events versus the background activity. This is a well-known technique in the literature (see \cite[Ch. 4.3]{Flandrin1999} for a hypothesis test approach to time-frequency detector design). An alternative approach is to track the variations of the auto-correlation function, estimated over sliding windows of the observations. If one adopts a model-based approach to auto-correlation estimation--- e.g. by fitting a time-varying auto-regressive (AR) model over sliding windows of the data--- temporal tracking of the model coefficients (or equivalently the root loci of the fitted model) is an efficient means for tracking the temporal nonstationarities \cite{kitagawa1985smoothness}. 

\subsection{Temporally Dynamic Sources and Innovations Monitoring}
Nonstationarity identification has also be performed in the time domain, by state-space modeling (\cite{Henrot2016,jamshidian2019temporally}). Suppose that the underlying multivariate signal of interest $\bm{s}_{k}$ is modeled by a dynamic system:
\begin{equation}
    \begin{array}{rl}
         \bm{s}_{k+1} =& \!\!\bm{f}_k(\bm{s}_{k}, \bm{w}_{k})\\
         \bm{x}_{k} =& \!\!\bm{g}_k(\bm{s}_{k}, \bm{v}_{k})
    \end{array}
\label{eq:basemodel}
\end{equation}
where $\bm{f}_k(\cdot,\cdot)$ and $\bm{g}_k(\cdot,\cdot)$ are generally nonlinear, differentiable and smooth functions considered as process and observation models, respectively. A very popular class of estimation schemes for such nonlinear models (including the Kalman or extended Kalman filter family) is in the form of a recursive estimator:
\begin{equation}
    \begin{array}{rl}
    \hat{\bm{s}}_{k}^+ & = \hat{\bm{s}}_{k}^- + \mathbf{K}_{k}\bm{i}_k, \quad \hat{\bm{s}}_{0}^+ = \bar{\bm{s}}_0\\
    \hat{\bm{s}}_{k+1}^- & = \bm{f}_k(\hat{\bm{s}}_{k}^+,\bar{\bm{w}}_k)
    \end{array}
\label{eq:FilterClass}
\end{equation}
where $\hat{\bm{s}}_{k}^-$ and $\hat{\bm{s}}_{k}^+$ denote the state vector estimates before and after using the $k$th observation, respectively; $\bm{i}_k\stackrel{\Delta}{=}\bm{x}_{k} - \bm{g}_k(\hat{\bm{s}}_{k}^-,\bar{\bm{v}}_k)$ is known as the \textit{innovation process}; and $\mathbf{K}_{k}$ is a gain matrix to be determined based on the sample-wise estimation qualities (covariance matrices). For a well-functioning filter, the innovation process is spectrally white (its auto-correlation is an impulse). In fact, the whiteness of the innovation process spectra is an indicator of ``optimal filtering'', showing that all the predictable parts of the observations have been retrieved, leaving only white noise behind. Now suppose that during the recursive time updates, the innovation process starts to deviate from white noise (becomes correlated in time). This behavior can be associated to changes in the underlying system's dynamics due to nonstationarity in the mixed signals or model parameters. This property was recently used in \cite{jamshidian2019temporally} to set up a hypothesis test for the noninvasive detection and extraction of fetal ECG.







\subsection{Challenges and Reservations}
\subsubsection{Insufficient Information}
\label{sec:cod}
Nonstationary events can suffer from the problem of insufficient information. In fact, for short nonstationary events, the sample estimates of the second or higher-order matrices/tensors are inaccurate. Therefore, model-based approaches are required to improve the statistics estimated from limited nonstationary points. Recent developments in \textit{random matrix theory} can also be used in this context, where methods have been developed that guarantee the accuracy of eigenvalue decomposition, despite the inaccuracies of sample covariance matrices \cite{engle2017large,tiomoko2019improved}. The extension of these techniques to AJD is currently an open problem.


\subsubsection{Rate of Change and Nonstationarity Transitions}
\label{sec:changerate}
Another point of reservation is the rate of signal attribute variations and the margins of nonstationary time windows. This problem has been addressed in the theory of \textit{abrupt change detection} \cite{basseville1993detection}, which is highly influential for the hereby studied problem.  

\section{Nonstationary Component Analysis (NSCA)}
\label{sec:NSCA}
To this point, we reviewed various methods for detecting nonstationary events. In this section, we study how the detected nonstationarities can be used during source separation, assuming a constant mixing matrix $\mathbf{A}_k=\mathbf{A}$.


\subsection{Single Hypothesis Tests and Generalized Eigenvalue Decomposition} 
Consider multivariate signals $\bm{x}_k \in \mathbb{R}^n$ ($k\in \mathcal{T}$), where $\mathcal{T}$ denotes the entire set of available samples and $\mathcal{P} \subset \mathcal{T}$ is a subset of the sample points considered as nonstationary or \textit{odd events}, with statistical properties that differ from the other samples. For this problem, the sample-wise hypothesis test in (\ref{eq:hypothesistestGeneral}) can be restated as follows: 
\begin{equation}
\begin{array}{cc}
         \mathcal{H}_0: k \notin \mathcal{P}, &
         \mathcal{H}_1: k \in \mathcal{P}
    \end{array}
\label{eq:hypotestrepeated}    
\end{equation}
The covariance matrices of the data over the samples of each hypothesis and the total data covariance matrix are defined as follows:
\begin{equation}
\begin{array}{rl}
         \mathbf{C}_{0} = \mathbb{E}_{k \notin \mathcal{P}}\{(\bm{x}_k-\mathbf{m}_{0})(\bm{x}_k-\mathbf{m}_{0})^T\} \\
         \mathbf{C}_{1} = \mathbb{E}_{k \in \mathcal{P}}\{(\bm{x}_k-\mathbf{m}_{1})(\bm{x}_k-\mathbf{m}_{1})^T\}\\
         \mathbf{C}_x = \mathbb{E}_{k \in \mathcal{T}}\{(\bm{x}_k-\mathbf{m}_{x})(\bm{x}_k-\mathbf{m}_{x})^T\}
    \end{array}
\label{eq:hypothesistestMatrices}
\end{equation}
where $\mathbb{E}\{\cdot\}$ denotes averaging over the indicated subset of samples, and $\mathbf{m}_{0} \stackrel{\Delta}{=} \mathbb{E}_{k \notin \mathcal{P}}\{\bm{x}_k\}$, $\mathbf{m}_{1} \stackrel{\Delta}{=} \mathbb{E}_{k \in \mathcal{P}}\{\bm{x}_k\}$ and $\mathbf{m}_{x} \stackrel{\Delta}{=} \mathbb{E}_{k}\{\bm{x}_k\}$ represent the class-wise and total sample means, respectively. Now, defining
\begin{equation}
	\bm{y}_k = \mathbf{W}^T\bm{x}_k
\label{eq:separatedsources}	
\end{equation}
where $\mathbf{W} \in \mathbb{R}^{n \times n}$, generalized eigenvalue decomposition (GEVD) of the matrix pair $(\mathbf{C}_0,\mathbf{C}_x)$ can be used to obtain $\mathbf{W}$, such that:
\begin{equation}
		\mathbf{W}^T \mathbf{C}_0 \mathbf{W} = \bm{\Lambda}, \quad \mathbf{W}^T \mathbf{C}_x \mathbf{W} = \mathbf{I}_n
\label{eq:EVD}		
\end{equation}
where $\bm{\Lambda} = \Diag(\lambda_1, \ldots, \lambda_n)$ is a diagonal matrix of generalized eigenvalues corresponding to the eigenmatrix $\mathbf{W}=[\mathbf{w}_1,\ldots,\mathbf{w}_n]$, with real eigenvalues sorted in ascending order on its diagonal. In addition, the first eigenvector $\mathbf{w}_1$, corresponding to the largest generalized eigenvalue $\lambda_1$, maximizes the \textit{Rayleigh quotient} \cite{Strang1988}:
\begin{equation}
J(\mathbf{w}) = \frac{\mathbf{w}^T \mathbf{C}_0 \mathbf{w}}{\mathbf{w}^T \mathbf{C}_x \mathbf{w}}.
\label{eq:Rayleigh}
\end{equation}
By using the Rayleigh quotient, it is straightforward to show that for single hypothesis test in (\ref{eq:hypotestrepeated}), the joint diagonalizer is exactly equivalent to the one obtained by GEVD of the matrix pair $(\mathbf{C}_1,\mathbf{C}_x)$, but this time sorting the eigenvalues in descending order. Another property of this transform is that $\mathbf{W}$ retrieves uncorrelated linear mixtures of $\bm{x}_k$ and $y_k=\mathbf{w}_1^T\bm{x}_k$ has maximal energy during the nonstationary time epochs. Therefore, $\mathbf{W}$ extracts the nonstationary events, corresponding to $\mathcal{H}_1$, from the other background signals.

\subsection{Multiple Hypothesis Tests and Approximate Joint Diagonalization of Eigenmatrices} 
\label{sec:multipletest}
The idea of sample clustering before source separation, can be extended to multiple classes, using a multiple hypothesis test. This idea is very close to a well-known method for separation of nonstationary sources, developed in \cite{pham2001blind}. 
In this case, the total data sample set $\mathcal{T}$ can be partitioned into $K$ classes/clusters: $\mathcal{T}=\mathcal{P}_1\cup\ldots \cup\mathcal{P}_K$, resulting in a multiple hypothesis test for the time samples:
\begin{equation}
 \mathcal{H}_i: k \in \mathcal{P}_i, \quad i = 1, \ldots, K
\label{eq:multiplehypothesistestGeneral}
\end{equation}
and the covariance matrices of each class is defined:
\begin{equation}
    \mathbf{C}_{i} \stackrel{\Delta}{=} \mathbb{E}_{k \in \mathcal{P}_i}\{(\bm{x}_k-\mathbf{m}_{i})(\bm{x}_k-\mathbf{m}_{i})^T\}
\label{eq:multihypothesistestMatrices}
\end{equation}
where $\mathbf{m}_{i} \stackrel{\Delta}{=} \mathbb{E}_{k \in \mathcal{P}_i}\{\bm{x}_k\}$. We may now seek the joint (approximate) diagonalizer $\mathbf{W} \in \mathbb{R}^{n \times n}$, such that
\begin{equation}
	\mathbf{W}^T \mathbf{C}_i \mathbf{W} = \bm{\Lambda}_i, \quad i = 1, \ldots, K
\label{eq:AJD}		
\end{equation}
are ``as diagonal as possible.'' It is known that for $K>2$, the diagonalization is only achieved approximately by using different variants of AJD. Depending on the application and diagonalization algorithm, the total covariance matrix $\mathbf{C}_x$, defined in (\ref{eq:hypothesistestMatrices}) may also be among the set of matrices to be diagonalized, to achieve decorrelated sources\footnote{Enforcing the diagonalization of $\mathbf{C}_x$ guarantees decorrelation of the extracted sources, at a cost of consuming $n(n-1)/2$ degrees of freedom of the matrix $\mathbf{W}$. This is why some BSS algorithms do not enforce whitening or sphering, but rather include the covariance matrix among the approximately diagonalized set of matrices, at a cost of reduced performance \cite{laheld1994adaptive}.}. 

\subsection{Source Separation vs Clustering/Classification}
\label{sec:sourceseparationversusclassification}
The problem of (approximate) joint diagonalization of a set of matrices is closely related to the problem of clustering and classification, which has readily been used in \textit{common spatial patterns} and similar algorithms \cite{koles1991quantitative,ramoser2000optimal,blankertz2007optimizing}. 
In the two-class (two-cluster) case, the relationship is explicitly evident from the maximization of the Rayleigh quotient in
(\ref{eq:Rayleigh}), which can be seen as the linear separator between two classes represented by their covariance matrices. The relationship in the multiple hypothesis test case (detailed in Section \ref{sec:multipletest}), can be non-rigorously explained as follows: by excluding an arbitrary subset of the sample points $\mathcal{P}_j$, one can set up a two class hypothesis test:
\begin{equation}
\begin{array}{cc}
    \mathcal{H}_0: k \notin \mathcal{P}_j, &
    \mathcal{H}_1: k \in \mathcal{P}_j
    \end{array}
\label{eq:hypothesistestoneversusall}
\end{equation}
which is a test of one versus the fusion of all other classes. The covariance matrix corresponding to the null hypothesis is: 
\begin{equation}
    \mathbf{C}_{\mathcal{T}\setminus \mathcal{P}_j} \stackrel{\Delta}{=}\displaystyle\sum_{i=1,\ldots, K, i \neq j} w_i \mathbf{C}_i
\end{equation}
where the differences of the cardinality of $\mathcal{P}_j$ and the significance of each matrix can be adjusted by the weights $w_i$. Now, suppose that the separation matrix $\mathbf{W}$ satisfies (\ref{eq:AJD}), i.e. using some typical measure of joint diagonality, $\bm{\Lambda}_i$ are ``as diagonal as possible'' \cite{ComonJutten2010handbook}. Therefore, the matrix $\mathbf{C}_{\mathcal{T}\setminus \mathcal{P}_j}$ is also ``approximately diagonalized'' by $\mathbf{W}$, and we have
\begin{equation}
\begin{array}{cc}
     \mathbf{W}^T \mathbf{C}_{\mathcal{T}\setminus \mathcal{P}_j} \mathbf{W} = \sum_{i \neq j} w_i \bm{\Lambda}_i, & 
     \mathbf{W}^T \mathbf{C}_j \mathbf{W} = \bm{\Lambda}_j
\end{array}
\label{eq:excludedsum}    
\end{equation}
Therefore, defining the vector of eigenvalue ratios
\begin{equation}
\mathbf{d}_j = \Diag[\bm{\Lambda}_j(\sum_{i \neq j}w_i\bm{\Lambda}_i)^{-1}], \quad j = 1, \ldots, K
\label{eq:eigratio}
\end{equation}
for each $j$, the eigenvector corresponding to the maximum entry of this vector, is ``approximately'' the separator of class samples $j$ versus the union of the other sample labels. In other words, if $m$ 
is the index of the maximum entry of the vector $\mathbf{d}_j$, then
$y_k=\mathbf{w}_m^T \bm{x}_k$ mostly resembles the samples of class $\mathcal{P}_j$. Since this property holds for all $j$, the separation method based on AJD can be seen as a method that maximizes the total between-class distances. However, the relationship between $m$ and $j$ is not necessarily one-to-one. In other words, for any class $j$, represented by its covariance matrix $\mathbf{C}_{j}$, there exists a separating vector $\mathbf{w}_m$ that maximally discriminates class $j$ from the union of others. But the same vector may happen to be the maximal separator for another class as well. Note also that the result of AJD is different from the solution obtained from the exact one versus the union of all other classes, which would be obtained by the GEVD of the matrix pair $(\mathbf{C}_{j},\mathbf{C}_{\mathcal{T}\setminus \mathcal{P}_j})$. The latter guarantees the maximal distance between class $j$ and the other classes; but does not attempt to maximize the inter-distance between the other classes.

The above property can be used for a ``two-round targeted source separation'', which first applies unsupervised source-separation algorithms such as JADE \cite{Cardoso1993}, to extract independent components ($n$ clusters). In the second round, any desired channel among the estimated independent components can be refined by performing a two-class NSCA between the desired channel versus the other components. From the clustering perspective, the first unsupervised ICA can be considered as $n$-group clustering, and the second round being a one vs $(n-1)$ classifier.



\section{Case Study}
\label{sec:casestudy}
\begin{figure}[tbh]
\centering
\begin{subfigure}[Raw signals]{\includegraphics[trim=1.5in 0.05in 1.2in 0.5in,clip,width=0.98\columnwidth]{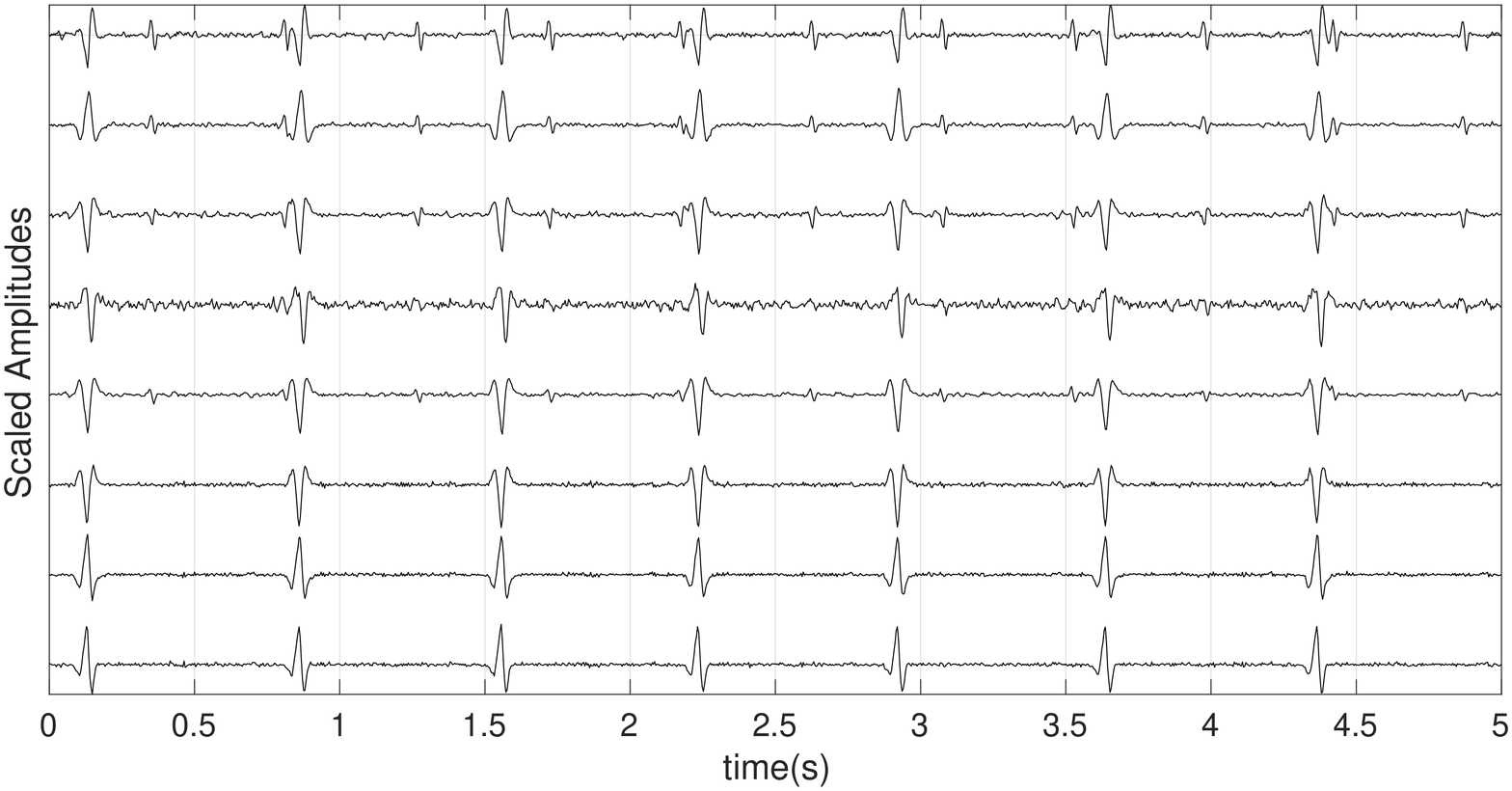}\label{fig:RawSignals}}
\end{subfigure}
\begin{subfigure}[Reference channel]{\includegraphics[trim=1.5in 0.04in 1.2in 0.1in,clip,width=0.98\columnwidth]{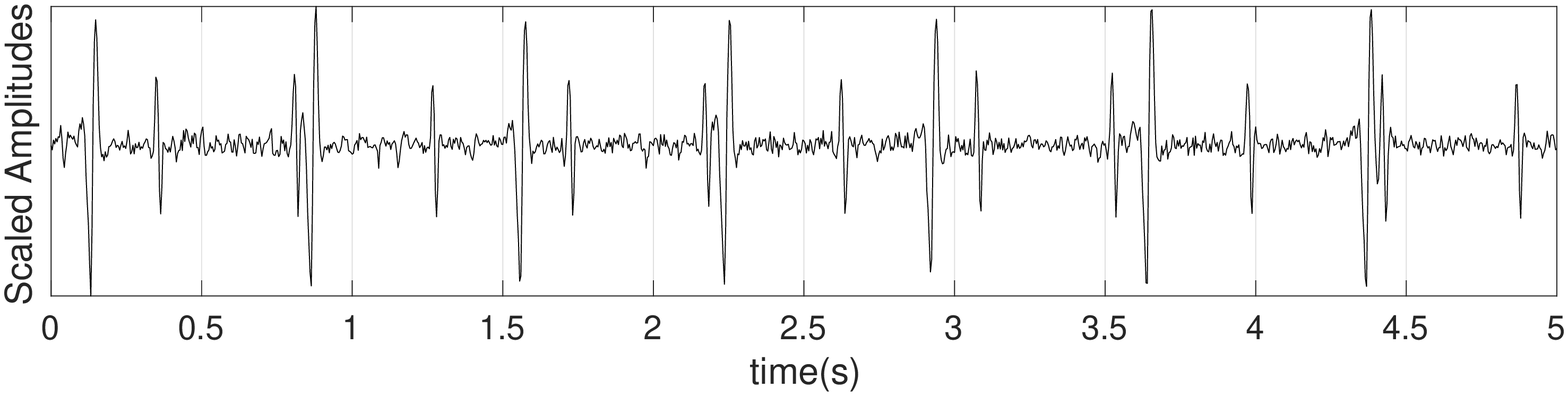}}
\end{subfigure}
\begin{subfigure}[Nonstationarity indexes]{\includegraphics[trim=1.5in 0.0in 1.2in 0.25in,clip,width=0.98\columnwidth]{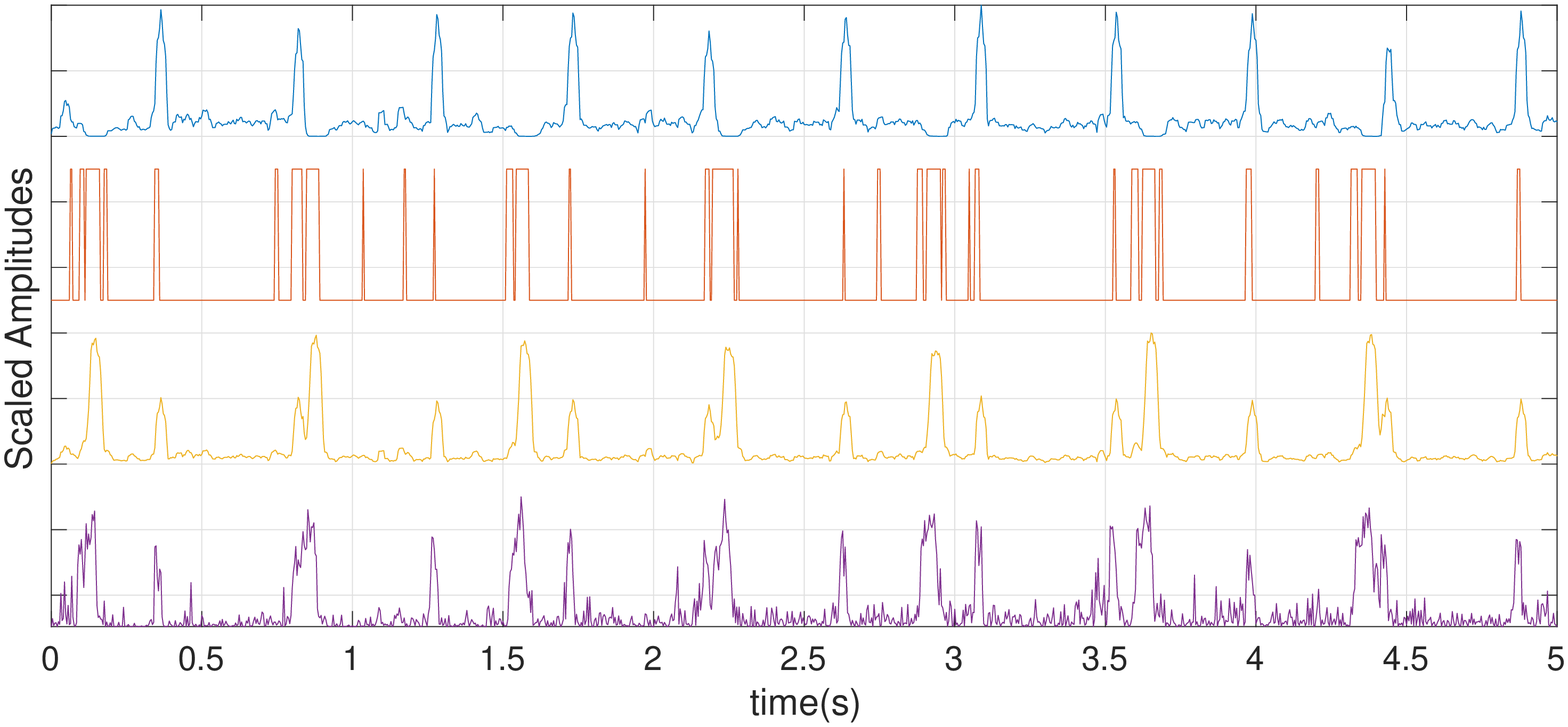}\label{fig:indexes}}
\end{subfigure}
\caption{(a) A sample fECG signal adopted the DaISy dataset; (b) First channel used as reference (c) Normalized nonstationarity indexes from top to bottom: innovation process, Anderson--Darling test, energy envelope and EASI nonlinear function norm}
\end{figure}

\begin{figure}[!ht]
\centering
\includegraphics[trim=1.5in 0.05in 1.2in 0.5in,clip,width=0.98\columnwidth]{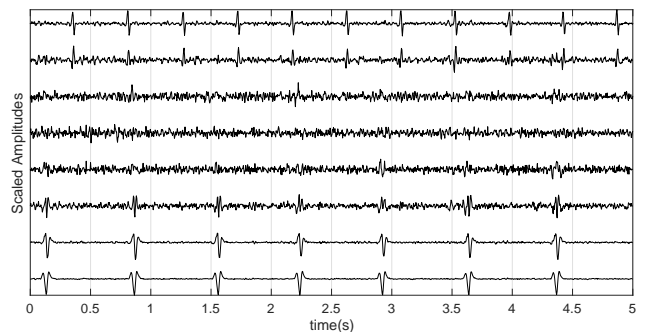}
\caption{The result of NSCA using the proposed innovations index}
\label{fig:NSCAResult}
\end{figure}

As a case study, we consider the problem of noninvasive fetal ECG (fECG) extraction from maternal abdominal recordings. In Fig.~\ref{fig:RawSignals}, a sample fECG signal from 5 maternal abdominal and 3 maternal chest leads is shown \cite{DeMoor}. In Fig.~\ref{fig:indexes}, the following four indexes introduced in Section \ref{sec:nonstationaritydetection} are shown for proof of concept: innovations index after maternal ECG (mECG) cancellation (cf. \cite{jamshidian2019temporally} for details), Anderson--Darling test with a Gaussian kernel fit, energy envelope, and EASI nonlinear function norm. The signals and indexes have been normalized for visual comparison. For this example, the innovation process index, has been the most discriminative, which is due to the fact that the innovation process after mECG cancellation mainly consists of the fECG and background noise. This index was next thresholded at a level of 50\% of its peak value to identify the nonstationary time epochs corresponding to the fECG. Fig.~\ref{fig:NSCAResult} demonstrates the results of NSCA by using this index, which shows the fECG in its first channels. The Matlab codes for generating these results are online available in the \textit{open-source electrophysiological toolbox (OSET)} \cite{OSET3.14}.


\section{Conclusion}
\label{sec:conclusion}
In this paper, various criteria which have been developed for the extraction of nonstationary components from multivariate data were reviewed from a unified hypothesis test perspective. It was shown how the problem is closely related to the joint (approximate) diagonalization of matrices/tensors containing second and higher-order statistics of the nonstationarty events and background stationary activities. The proposed approach was also shown to have a clustering/classification interpretation. Various aspects of this study can be extended in future research, including: 1) integration with deflation algorithms such as \cite{SJS2010}, for low-rank and degenerate mixtures; 2) using developments in random matrix theory to overcome the problem of low number of samples during nonstationary epochs of the data \cite{engle2017large,tiomoko2019improved}; 3) explicit integration of NSCA with classification and clustering algorithms.  

\bibliographystyle{IEEEtran}
\bibliography{IEEEabrv,References}
\end{document}